\documentclass[aps,pra,twocolumn,longbibliography,footinbib,groupedaddress]{revtex4-1}
\usepackage{graphicx}
\usepackage{multirow}
\usepackage{amsmath}
\usepackage{amssymb}
\usepackage[utf8]{inputenc}
\usepackage{adjustbox}
\usepackage{booktabs}

\newcommand {\gtau} {$g^{(2)}(\tau)$ }
\newcommand {\gr} {$g^{(2)}(r)$ }
\newcommand {\gtaur} {$g^{(2)}(\tau,r)$ }
\newcommand {\el} {\mathrm{el}}
\newcommand {\PCyg} {P~Cyg}

\newcommand {\aap} {A\&A }

\newcommand \dd[1] { \,\textrm d{#1}}   % infintesimal

\providecommand{\e}[1]{\ensuremath{\times 10^{#1}}}
\DeclareRobustCommand{\ion}[2]{\textup{#1\,\textsc{\lowercase{#2}}}}

\begin{document}

\title{Intensity interferometry of P Cygni in the H$\alpha$ emission line: towards distance calibration of LBV supergiant stars}

\author{J.-P. Rivet$^{1}$}\email{Jean-Pierre.Rivet@oca.eu}
\author{A.~Siciak$^{2}$}
\author{E.~S.~G.~de~Almeida$^{1}$}
\author{F.~Vakili$^{1,3}$}
\author{A.~Domiciano~de~Souza$^{1}$}
\author{M.~Fouch\'e$^{2}$}
\author{O.~Lai$^{1}$}
\author{D.~Vernet$^{4}$}
\author{R.~Kaiser$^{2}$}
\author{W.~Guerin$^{2}$}\email{William.Guerin@inphyni.cnrs.fr}

\affiliation{$^{1}$Universit{\'e} C{\^o}te d'Azur, Observatoire de la C{\^o}te d'Azur, CNRS, Laboratoire Lagrange, France}
\affiliation{$^{2}$Universit{\'e} C{\^o}te d'Azur, CNRS, Institut de Physique de Nice, Valbonne, France}
\affiliation{$^{3}$Department of Physics, Shahid Beheshti University, G.C., Tehran, Iran}
\affiliation{$^{4}$Universit{\'e} C{\^o}te d'Azur, Observatoire de la C{\^o}te d'Azur, CNRS, UMS Galil\'ee, Nice, France}

\begin{abstract}
We present intensity interferometry of the luminous blue variable \PCyg\ in the light of its H$\alpha$ emission performed with 1\,m-class telescopes. We compare the measured visibility points to synthesized interferometric data based on the CMFGEN physical modeling of a high-resolution spectrum of \PCyg\ recorded almost simultaneously with our interferometry data. Tuning the stellar parameters of \PCyg\ and its H$\alpha$ linear diameter we estimate the distance of \PCyg\ as $1.56\pm0.25$~kpc, which is compatible within $1\sigma$ with $1.36\pm0.24$~kpc reported by the Gaia DR2 catalogue of parallaxes recently published. Both values are significantly smaller than the canonic value of $1.80\pm0.10$~kpc usually adopted in literature. Our method used to calibrate the distance of \PCyg\ can apply to very massive and luminous stars both in our galaxy and neighbour galaxies and can improve the so-called Wind-Momentum Luminosity relation that potentially applies to calibrate cosmological candles in the local Universe.
\end{abstract}

\maketitle

%%%%%%%%%%%%%%%%% BODY OF PAPER %%%%%%%%%%%%%%%%%%

%%%%%%%%%%%%%%%%%%%%%%%%%%%%%%%%%%%%%%%%%%%%%%%%%%%%%%%%%%%%%
%%%%%%%%%%%%%%%%%%%%%%%%%%%%%%%%%%%%%%%%%%%%%%%%%%%%%%%%%%%%%
\section{Introduction}\label{sec.intro}

%The purpose of this paper is two-fold. On the one hand we demonstrate that some of the limitations of intensity interferometry (hereafter II), as it was pioneered by Hanbury Brown and his team more than 50 years ago~\cite{HBT:1956,HBT:1968} can be overcome, opening new paths in observational astronomy. This has been achieved by borrowing modern photonics components from the quantum optics community, that we set-up on rather modest optical telescopes of $1$~m class~\cite{Guerin:2017,Guerin:2018,Rivet:2018,Lai:2018}. On the other hand we show that, even with its remaining present limitations, II can provide new and useful information on the fundamental stellar parameters and the mechanisms that govern the physics of massive stars and their mass loss\,: more precisely on the Luminous blue variable (LBV) archetype star \PCyg\ (HD193237)~\cite{Najarro:2001}

% slightly shorter version
The purpose of this paper is to show that, even with its present limitations, intensity interferometry can provide new and useful information on the fundamental stellar parameters and the mechanisms that govern the physics of massive stars and their mass loss\,: more precisely on the Luminous blue variable (LBV) archetype star \PCyg\ (HD193237)~\cite{Najarro:2001}.

Intensity interferometry (hereafter II), as imagined by Hanbury Brown and Twiss in the
1950's \cite{HBT:1956}, culminated in the early 1970s by providing the first
systematic catalogue of the angular diameter (in the visible) of 32 stars observed with the
Narrabri $200$\,m-baseline interferometer~\cite{HBT:1974}. In addition to this, Hanbury Brown
and his team explored different phenomenological effects, such as flattening of rapidly rotating stars, close binary stars and their parameters~\cite{HBT:1971}, scattering effects occurring in the massive wind of blue supergiants and emission carbon line extent of a Wolf-Rayet star~\cite{HBT:1970}. An extensive review of these experiments is described by Hanbury Brown in his book on the Narrabri interferometer~\cite{HBT:book}, which stopped operating in the early seventies. More contemporary, Cherenkov arrays of telescopes have been considered to revive II with much larger collectors in size and much longer baselines, aiming at stellar surface imaging by aperture synthesis interferometry on a much broader class of targets~\cite{Dravins:2016,Kieda:2019,Kieda:2019b}. Successful demonstrations of II with Cherenkov telescopes have been reported very recently~\cite{Matthews:2019,Acciari:2020}. In this context our group started a number of pilot experiments in 2016 using two modest 1\,m size optical telescopes. After the successful observations of temporal and spatial bunching on a few bright stars at $780$~nm~\cite{Guerin:2017,Guerin:2018,Rivet:2018,Lai:2018}, we decided to observe emission-line stars. The LBV star \PCyg\ is a very good candidate due to its strong H and He emission lines.

In the following we shortly discuss the photometric and spectral variability of \PCyg\, especially for its H$\alpha$ line that is relevant to our II observations. We then summarize the 3 long-baseline interferometric studies available in the literature that shed light on the present results.

Along with $\eta$ Car, \PCyg\ is the brightest LBV star in the sky, having undergone a giant eruption in the 17th century and for which evolutionary change has been recorded from its apparent magnitude by ~\cite{Lamers:1992} over 3 centuries. More recently \PCyg\ was studied by~\cite{Markova:2001} using U,B,V photometry and H$\alpha$ emission, including equivalent width (EW) monitoring over 13.8 years from 1985 to 1999. These authors find that \PCyg\ undergoes a slow 7.4 year variation in its V magnitude, where the star becomes redder when it brightens, and vice versa. They also show that the H$\alpha$ EW changes in correlation with the photometric trend and conclude that the variable wind increases the photospheric radius while the effective temperature is decreasing. \cite{Markova:2001} concluded that the wind mass-loss rate of \PCyg\  increased of $\sim$19\% over a period of about 7 years. This increase in mass-loss rate implies an apparent stellar radius (pseudo-photosphere) larger by $\sim$7\%. Thus, angular diameter observations of this star also need to be monitored by spectrometry of the H$\alpha$ line and simultaneous photometry~\cite{pollmann:2013} in order to correctly analyze and interpret the interferometric data. Our present work meets the H$\alpha$ spectroscopy criterion.

Milli-arc-second (mas) resolution observations of \PCyg\ trace back to GI2T spectrally-resolved interferometry based on visibility and differential phase of the H$\alpha$ emission line~\cite{Vakili:1997}. These quantities were determined as a function of the Doppler-shift across the H$\alpha$ line profile and gave the first angular diameter of
\PCyg's envelope as well as a limit to its extent in HeI line. In addition, the signature of an asymmetry in the wind of \PCyg\ was concluded from a differential phase occurring at the blue absorption component of the H$\alpha$ line. It is worth noting that \cite{Vakili:1997} estimated the diameter of \PCyg\ in H$\alpha$ as $5.52 \pm 0.47$\,mas assuming a simple equivalent uniform disc, without separating the star photosphere and its envelope emission. This single shot observation and study of \PCyg\ was followed in 1997 by adaptive optics imaging in the H$\alpha$ line through a 1\,nm filter and in its continuum vicinity with a 1.5\,m telescope, corresponding to 0.1'' diffraction limit resolution~\cite{Chesneau:2000}. This adaptive optics imaging aimed at first to determine the large scale extent of \PCyg's envelope as it had been previously witnessed by HST observations~\cite{Nota:1995}, and secondly detect, if possible, the propagation after 4 years of the heterogeneities of \PCyg's wind detected by the GI2T. The latter expectation was roughly confirmed whilst it was clearly confirmed also that any high angular resolution observation of \PCyg\ should consider the central LBV engine, its mass-loss envelope out to thousands of stellar radii, even though dilution factor would make this a high-contrast imaging challenge.

\PCyg\ was then observed between 2005 and 2008 with the NPOI interferometer~\cite{Balan:2010} with simultaneous spectroscopy to relate any angular diameter variation with the H$\alpha$ line profile and/or emission strength. These observations used a much broader filter than the above-mentioned studies and modeling of the envelope was conducted for the equivalent H$\alpha$ width emission using different circular shapes. Finally the authors concluded that the data are best fitted with a double Gaussian structure of $5.64 \pm 0.21$ and $1.80 \pm 0.13$\,mas for \PCyg's envelope. In addition they found no asymmetry of the envelope and less than 10\% variation in size between 2005 and 2008. To make NPOI results comparable to previous GI2T measures, \cite{Balan:2010} considered also the simple model of a uniform disc for \PCyg's emission envelope including indifferently the photosphere as well as its envelope. They found uniform disk angular diameters ranging from 8.4 to 10.2 mas on the seasonal observations between 2005 and 2008, a result that significantly differs from those by the GI2T single baseline data. Balan~\textit{et al.} finally concluded that this discrepancy might result from photospheric flux variability and opacity changes through the multiple wind layers of \PCyg.

More recent long baseline interferometry of \PCyg, covering the period of 2006 to 2010, has been reported by \cite{Richardson:2013} using the CHARA interferometer at Mount Wilson. These observations were accompanied by simultaneous infrared (IR) photometry and spectroscopy to monitor any change in the angular size of \PCyg\ related to the activity at the base of the wind and its impact on eventual fine structures within the mass loss. These observations differ from previous studies since they have been performed in the IR at $1.6$~$\mu$m (H band), but can still compare to GI2T and NPOI observations. A first important issue of CHARA-MIRC conclusions consists on setting an angular diameter of $0.96 \pm 0.02$~mas for the wind component of \PCyg\ at its photospheric base with about $45\%$ of the H-band flux. This angular diameter is significantly larger than the $0.41$~mas~\cite{Najarro:1997} that was adopted for \PCyg\ photospheric diameter used by the GI2T paper for instance~\cite{Vakili:1997}. Additionally, multiple baseline performed with CHARA-MIRC at two epochs in August~2010 and September~2011 were used by \cite{Richardson:2013} to reconstruct an image of \PCyg\ from Earth rotation synthesis data. Furthermore, these authors used the non-LTE radiative transfer code CMFGEN (spherically symmetric wind) to compare the observed visibility curve of \PCyg\ to the predicted one. Whilst no significant departure was found from circular symmetry, Richardson~\textit{et al.} concluded that \PCyg\ is best explained  by a two component model consisting of a uniform disk photosphere unresolved by CHARA at its 0.56\,mas resolution in the H band and a $0.96 \pm 0.02$\,mas Gaussian halo emitted from the inner regions of the stellar wind of \PCyg. The difference between this result and the $5.5$~mas size found by \cite{Vakili:1997} can be explained by a larger wind-emitting volume because of its higher optical depth in H$\alpha$ diameter. Besides these interferometric studies, such a spherically symmetric wind around \PCyg\ is also supported by H$\alpha$ spectroscopy \cite{Richardson:2011}.

Due to their limited spatial frequency content, interferometric observations require a model for their interpretation and as shown by the review of these high-resolution results, models that introduce the least amount of a priori information (e.g. uniform disk or Gaussian profile) are usually chosen, yielding limited information, such as the apparent diameter. For our II campaign reported herein, we chose an alternative approach, using the best physical parameters of \PCyg\ from the CMFGEN code~\cite{Hillier:1998} that reproduce high resolution spectrometry of the star obtained quasi-simultaneously to our II campaign, to compute the intensity distribution (and its associated visibility) of \PCyg\ in the emission line, which can be directly compared to our measured visibilities. This additional information constrains the physical size of \PCyg\ and allows us to estimate the only remaining free parameter, which is its distance. We believe that the association of physical modeling of stellar parameters of LBVs from spectroscopy with interferometric  observations has the potential to be a powerful method to refine the first few rungs of the cosmological distance ladder.

The paper is organized as follows. In the next section we describe our experimental setup and the observing conditions. In section \ref{sec.single-telescope} we show the results of our single-telescope observations, which can be used to calibrate the visibility at zero baseline. This also corresponds to measuring the \emph{temporal} intensity correlation, related to the width of the spectral line. Then in section~\ref{sec.two-telescope} we present the \emph{spatial} intensity correlation measurements performed with two telescopes separated by 15\,m. We observe a reduction of the contrast of the correlation, demonstrating a partial resolution of the emitting envelope. Finally in section~\ref{sec.Armando} we present our CMFGEN best model to compare the expected and the measured visibilities using the star distance as the only free parameter. We then conclude and draw some perspectives.

%%%%%%%%%%%%%%%%%%%%%%%%%%%%%%%%%%%%%%%%%%%%%%%%%%%%%%%%%%%%%
\section{Experimental setup}\label{sec.etup}

\subsection{Principle}\label{sec.principle}

Stellar intensity interferometry is based on the measurement of the temporal and spatial correlations between the fluctuations of light collected by two telescopes distant by $r$. The quantity of interest is the intensity correlation function given by
\begin{equation}
g^{(2)}(r, \tau) = \frac {\left\langle I(t, 0)I(t+\tau, r)\right\rangle}{\left\langle I(t, 0)\right\rangle \left\langle I(t, r)\right\rangle},
\end{equation}
where the brackets denote the average over time $t$. For a classical (non-quantum) source of light, correlations are maximum at zero delay ($\tau=0$) and zero separation ($r=0$). For a ``chaotic'' (incoherent) source, Gaussian statistics on the field fluctuations leads to $g^{(2)}(0,0)=2$. On the contrary, at large separation $r$ and delay $\tau$, the fluctuations become uncorrelated and the \gtaur tends to 1. The decrease from 2 to 1 of the $g^{(2)}$ function with the time delay $\tau$ is related to the temporal coherence time $\tau_\mathrm{c}$, which is inversely proportional to the spectrum width. The decrease of the $g^{(2)}$ function with the separation $r$ is related to the spatial coherence of the source, i.e. the usual ``visibility'' $V(r)$ measured in direct (amplitude) stellar interferometry~\cite{Labeyrie:book}.

We can therefore use the following equation, valid for chaotic light~\cite{Loudon:book},
\begin{equation}\label{eq.Siegert}
g^{(2)} (r, \tau) = 1 + |V(r)|^2 |g^{(1)} (\tau)|^2,
\end{equation}
where $g^{(1)}(\tau)$ is the first order (field-field) temporal correlation function, related to the optical spectrum $S(\omega)$ by a Fourier transform (Wiener-Khinchin theorem),
\begin{equation}\label{eq.WienerKhinchin}
S(\omega) \propto \mathfrak{F}[g^{(1)}(\tau)].
\end{equation}
Similarly, the visibility is related to the brightness distribution of the source by a Fourier transform, like in direct interferometry. Note that we suppose here that the detected light is polarized, otherwise it amounts at reducing the visibility by a factor 2.

In practice, the coherence time $\tau_\mathrm{c}$, which gives the width of the \gtau function, is often too short to be resolved by the electronic detection chain, whose finite timing resolution $\tau_\el$ introduces an uncertainty on the arrival time of each photon. In that case the measured ``bunching peak'' \gtau has a width given by $\tau_\el \gg \tau_\mathrm{c}$, and a height, which we call the contrast $C = g^{(2)} (0, 0)-1$, reduced from $C = 1$ to $C \sim \tau_\mathrm{c}/\tau_\el \ll 1$. This contrast has to be calibrated as it corresponds to the maximum (zero-baseline) squared visibility.

\subsection{Instrumental setup}\label{sec.sub_setup}

Our experimental setup has been described in detail in previous publications~\cite{Guerin:2017,Guerin:2018}. In short, it consists first in a coupling assembly (CA) set at the focus of the telescope, which allows injecting light into a multimode fiber (MMF) of diameter 100\,$\mu$m. Then, we use single-photon avalanche photodiodes (SPADs) to detect light in the photon-counting regime and digital electronics in order to compute the $g^{(2)}$ function. The SPADs have a timing jitter of $\simeq 450$\,ps each, which gives a temporal resolution $\tau_\el \simeq \sqrt{2}\times450 \simeq 640$\,ps for the correlation function.

\begin{figure}[b]
    \centering\includegraphics[width=\columnwidth]{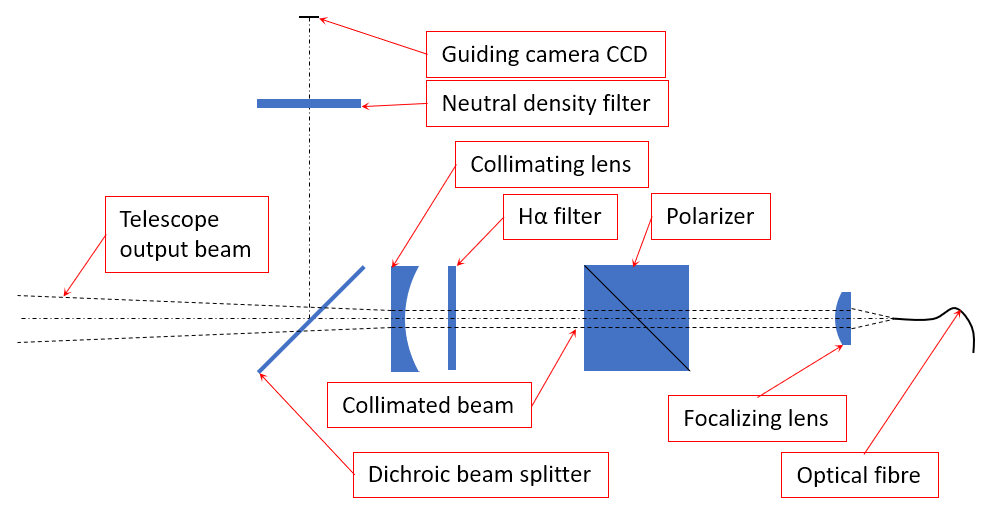}%
    \caption{Scheme of the coupling assembly set at the telescope focus in order to perform spectral and polarization filtering, and injection into a multimode fiber, which transports the light to the detection chain. The H$\alpha$ $1$~nm filter operates in nominal conditions, {\em i.e.} on a collimated beam.
    }
    \label{fig.Setup}
\end{figure}

Compared to our previous experiments\,\cite{Guerin:2017,Guerin:2018}, we have modified the CA in order to collimate the optical beam before its transmission through the filter, in order to have a more precise control on the filter width and central wavelength. This is indeed more critical when one wants to select a specific spectral line. The new CA is described in Fig.~\ref{fig.Setup}. As previously, there is first a dichroic mirror that reflects part of the light to a guiding camera. The transmitted light is then collimated by a diverging lens (focal length $f = -50$\,mm ). The collimated beam goes through a filter of width $\Delta\lambda = 1$\,nm (FWHM), centered at $\lambda = 656.3$\,nm (H$\alpha$ line), with a peak transmission of 95\%, and then to a polarizer, before being focused by a converging lens ($f = 20$\,mm) on the fiber tip.

The observations have been performed at the C2PU facility at the Plateau de Calern site of Observatoire de la C{\^o}te d'Azur (OCA). The two, quasi-identical, telescopes have a diameter of $1.04$\,m with an F$/12.5$ aperture in a Cassegrain configuration. With the CA, the total equivalent focal length is $5.2$\,m. The two CAs are identical and we have checked, using an artificial star in the lab, that they produce identical correlation functions.

\subsection{Observation conditions}\label{sec.observing}

The observations of \PCyg\ were performed in August 2018 over 8 nights. The main characteristics of the observing runs are summarized in Table~\ref{tab.ObsRuns}.

\begin{table*}
\centering
\caption{Main circumstances for the observing runs performed on P~Cyg over 8 nights. ``Configuration'' indicates the performed experiment, either $g^{(2)}(\tau)$ (single-telescope experiment) of $g^{(2)}(r)$ (two-telescope experiment). Begin and end dates are in UTC (ISO~8601 compact format). $a$ is the air mass range. The seeing information is provided by the GDIMM instrument~\cite{Ziad:2012,Aristidi:2014}
of the CATS station (Calern Atmospheric Turbulence Station)~\cite{Chabe:2016}
. The numbers are given as median values over the whole nights.}
\label{tab.ObsRuns}
\renewcommand{\tabcolsep}{2pt}
\begin{tabular}{lcccc}
\hline
Configuration & Begin & End & $a$ & Seeing\\                                                           %     date ,      heure de la montre
\hline
\,\gtau\, & \,$20180801T2102Z$\, & \,$20180802T0111Z$\, & \,$1.00 \rightarrow 1.10$\, & \,$1.29''$\,\\    % 1  - 2  August, 23h02 - 3h11
\,\gtau\, & \,$20180802T2025Z$\, & \,$20180803T0154Z$\, & \,$1.00 \rightarrow 1.18$\, & \,$0.66''$\,\\    % 2  - 3  August, 22h25 - 3h54
\,\gtau\, & \,$20180804T0040Z$\, & \,$20180804T0309Z$\, & \,$1.06 \rightarrow 1.44$\, & \,$1.10''$\,\\    % 3  - 4  August, 2h40 - 5h09
\,\gtau\, & \,$20180806T1943Z$\, & \,$20180806T2205Z$\, & \,$1.02 \rightarrow 1.22$\, & \,$0.56''$\,\\    % 6  - 7  August, 21h43 - 0h05
\,\gr \,  & \,$20180807T0054Z$\, & \,$20180807T0356Z$\, & \,$1.10 \rightarrow 1.79$\, & \,$0.56''$\,\\    % 6  - 7  August, 2h54 - 5h56
\,\gr \,  & \,$20180807T2212Z$\, & \,$20180808T0353Z$\, & \,$1.00 \rightarrow 1.80$\, & \,$0.60''$\,\\    % 7  - 8  August, 0h12 - 5h53
\,\gr \,  & \,$20180808T2011Z$\, & \,$20180809T0350Z$\, & \,$1.00 \rightarrow 1.81$\, & \,$0.74''$\,\\    % 8  - 9  August, 22h11 - 5h50
\,\gr \,  & \,$20180809T2311Z$\, & \,$20180810T0332Z$\, & \,$1.01 \rightarrow 1.71$\, & \,n.a.    \,\\    % 9  - 10 August, 1h11 - 5h3
\,\gr \,  & \,$20180810T1940Z$\, & \,$20180811T0327Z$\, & \,$1.00 \rightarrow 1.70$\, & \,$1.19''$\,\\    % 10 - 11 August, 21h40 - 5h37
\hline
\end{tabular}
\end{table*}

The observing time was used in two configurations. In the first one, we used only one telescope (always the same), in order to measure the temporal intensity correlation function \gtau at zero baseline, as in \cite{Guerin:2017}. In principle, the contrast of the correlation function allows calibrating the visibility measured with two telescopes, which is the second configuration we used, as in \cite{Guerin:2018}.

%%%%%%%%%%%%%%%%%%%%%%%%%%%%%%%%%%%%%%%%%%%%%%%%%%%%%%%%%%%%%
\section{Temporal intensity correlation}\label{sec.single-telescope}

Performing intensity interferometry on an emission line puts an important constraint on the measurement procedure. Indeed, since the $g^{(2)}$ function depends on the spectrum, it is not possible to use a distant, unresolved star as calibrator for the visibility measurement, because this calibrator would have a different spectrum from the science target. For the same reason, it is not possible to calibrate the visibility with an artificial star in the laboratory, as in \cite{Guerin:2018}. As a consequence, there are two possibilities. The first is to measure the actual spectrum, use Eqs.~(\ref{eq.Siegert}-\ref{eq.WienerKhinchin}) and, knowing the temporal resolution of the detection chain, infer the expected bunching contrast for maximum visibility. The second is to perform a temporal intensity correlation measurement with a single telescope, as in \cite{Guerin:2017}, which serves as the zero-baseline visibility calibration. We do both in the following.

\subsection{H$\alpha$ spectrum of \PCyg\ and expected temporal correlation}\label{sec.spectrum}

Thanks to its strong H$\alpha$ emission line, \PCyg\ is a classical target for amateur spectroscopy, which enabled us to obtain a spectrum recorded only a few days after
our observations in the ARAS spectral database~\cite{ARAS}. This spectrum (resolution\,: $9000$, $4053$\,{\AA}$<\lambda<7763$\,{\AA}) was recorded by J.~Guarro i Fl{\'o} on August 14th, 2018.

\begin{figure}
    \centering\includegraphics{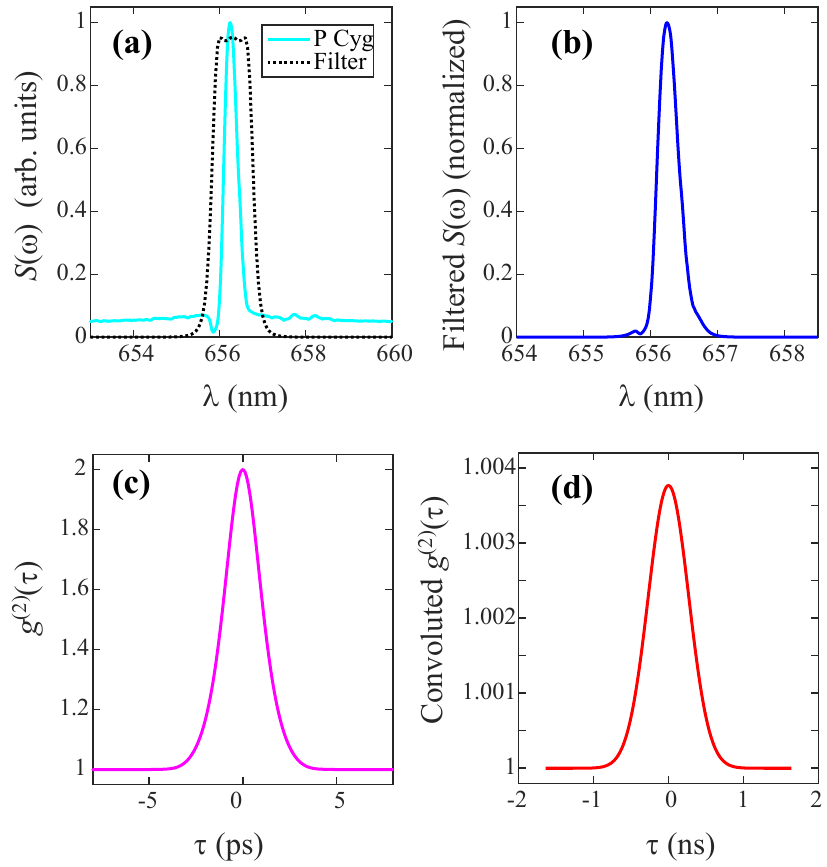}
    \caption{Spectrum and \gtau function. (a) Spectrum of \PCyg\ zoomed-in on the
    H$\alpha$ emission line and transmission spectrum $T$ of the filter (simulations
    provided by the manufacturer). (b) Filtered spectrum computed by multiplying the
    spectra of the star and of the filter. (c) Computed \gtau from the filtered
    spectrum using Eqs.~(\ref{eq.Siegert}-\ref{eq.WienerKhinchin}) and supposing
    maximum visibility. (d) Convoluted \gtau with the timing resolution of our
    acquisition chain. We have taken a Gaussian jitter of $450$~ps (FWHM) per detector.}
    \label{fig.spectrum}
\end{figure}

We show in Fig.~\ref{fig.spectrum}(a) the measured spectrum centered on the H$\alpha$ line, as well as the transmission spectrum of the $1$~nm filter set in the CA, as provided by the manufacturer. Multiplying the two spectra, we obtain the spectrum of the detected light [Fig.~\ref{fig.spectrum}(b)]. Note that at this scale, the variation with the wavelength of the other elements (reflectivity of mirrors, transmission of the atmosphere and of the dichroic plate, quantum efficiency of the detectors) is negligible.

From the filtered spectrum, one can numerically compute the \gtau function by using Eqs.~(\ref{eq.Siegert},\ref{eq.WienerKhinchin}). This theoretical \gtau function has a 100\% contrast and a width on the order of the picosecond [Fig.~\ref{fig.spectrum}(c)]. Experimentally we measure this function convolved by the response of the instrument, dominated by the jitter of the SPADs. The resulting expected \gtau function is depicted in Fig.~\ref{fig.spectrum}(d). Note the change of scales compared to Fig.~\ref{fig.spectrum}(c). The expected contrast is now $C_0 = 3.8\times 10^{-3}$.

\subsection{Measured temporal correlation}\label{sec.results-single}

We present in this section the measurement of \gtau with a single telescope observing
\PCyg. In this configuration the flux collected by the telescope is separated into two SPADs in order to overcome the dead time of the detectors~\cite{Guerin:2017}. This leads to some spurious correlations due to optical and electronic cross-talk between the detectors. These spurious correlations have to be characterized with a white source, for which the expected \gtau function is flat, and then removed from the signal~\cite{Guerin:2017}. The ``white'' signal has been measured in the lab after the observing run with a similar count rate.

\begin{figure}
    \centering\includegraphics{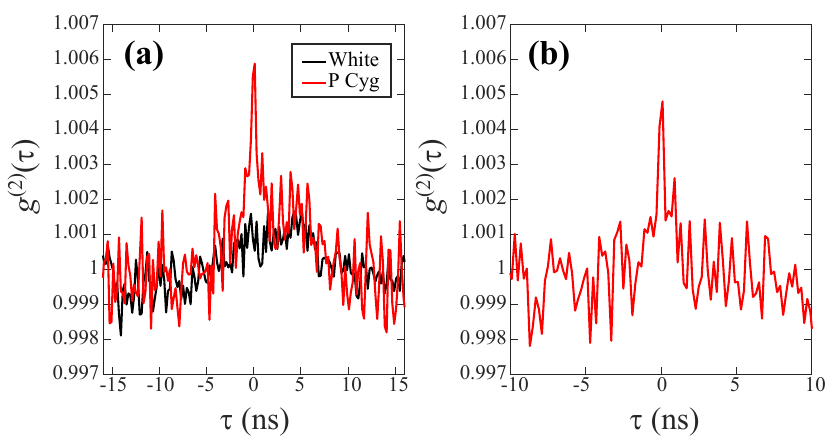}
    \caption{Measured temporal intensity correlation function. (a) Signal on the star and ``white'' acquired in the lab. (b) \gtau after removing the spurious correlations. The binning is 200\,ps.}
    \label{fig.g2tau}
\end{figure}

The count rate was in average $3.8\times 10^5$ counts per second (hereafter cps) per detector. The total observation time on \PCyg\ was $14$~hours over $4$~nights (Tab.~\ref{tab.ObsRuns}). The obtained \gtau functions are shown in Fig.~\ref{fig.g2tau}, with the direct measurements (\PCyg\ and ``white'') in panel (a), and the corrected
correlation function (after division by the ``white'' signal to remove the spurious
correlations) in panel (b).

The height of the bunching peak, defined as the maximum of the peak, is $C = (4.8 \pm 0.9)\times 10^{-3}$, in fair agreement with the expectation [Fig.\,\ref{fig.spectrum}(d)] given the uncertainty, estimated from the {\it rms} noise in the flat areas of the \gtau function. Note that a small systematic effect may also be present due to an imperfect removing of spurious correlations and explain a slightly higher value than expected ($C_0 = 3.8\times 10^{-3}$). For this reason, in section~\ref{sec.visibilities}, we will use the computed $C_0$ value to normalize the visibility data.

Besides providing a zero-baseline calibration for the spatial correlation measurement detailed in the next section, another important aspect of this temporal correlation experiment on an emission line is that the resulting \gtau function provides information on the emission line itself via the contrast $C \sim \tau_\mathrm{c}/\tau_\el$. Knowing the response function of the instrument (and thus $\tau_\el$) and with some assumption on the shape of the line, we can deduce the width $\Delta\lambda$ of the emission line via $\tau_\mathrm{c} = \lambda_0^2/c\Delta\lambda$. Here, approximating the line shape by a Gaussian, the temporal correlation measurement corresponds to a line width of FWHM $\Delta\lambda \simeq 0.3$\,nm, in agreement with the actual spectrum. Note also that the contrast of the \gtau function measured here is significantly higher than what it would be if it were determined by the 1-nm filter (the contrast would be $\sim 1.4\times10^{-3}$), which would be the case in the continuum~\cite{Guerin:2017}. This ``intensity-correlation spectroscopy'' technique~\cite{Goldberger:1966,Phillips:1967,Tan:2017} would be relevant for exotic, very narrow lines, that would be hard to characterize with standard spectroscopic techniques. With intensity correlation, the narrower the line, the higher the contrast.

%%%%%%%%%%%%%%%%%%%%%%%%%%%%%%%%%%%%%%%%%%%%%%%%%%%%%%%%%%%%%
\section{Spatial intensity correlation}\label{sec.two-telescope}

We now turn to the spatial correlation experiment, performed with two nearly-identical telescopes separated by 15\,m on an East-West basis~\cite{Guerin:2018}. The flux collected at each telescope is filtered and coupled to the MMF with an identical CA and detected by a SPAD. The count rate per detector was in average $8.8\times10^5$\,cps with a total acquisition time of 27 hours over 5 nights (Tab.~\ref{tab.ObsRuns}).

\begin{table*}
    \centering
    \caption{Summary of the observation results. $r$ is the average projected baseline (its uncertainty is the rms width of the baseline distribution), $T$ is the total integration time, $F$ is the detected count rate per detector averaged over the total integration time. It is roughly twice lower in the single-telescope experiment because the flux has to be divided into two detectors. The contrast $C = g^{(2)}(0)-1$ is the value of the correlation at zero delay given by the amplitude of the bunching peak, its uncertainty is the rms noise on the data. The two last columns correspond to the two possible normalization methods ($C_0 = 3.8 \times 10^{-3}$ is the zero-baseline contrast expected from the measured spectrum).}
    \label{tab.Results}
    \begin{tabular}{lrccccc}
        \hline
        $r$ (m) &  $T$ (h) & $F$ ($\times 10^3$ cps) & $C(r)$ $(\times 10^{-3})$ & $C(r)/C(0)$ & $C(r)/C_0$ \\
        \hline
        $0$ &  14.5  & 380 & $4.80 \pm 0.93$ &  1 & $1.26 \pm 0.24$ \\
        $10.7 \pm 0.7$ &  8    & 826 & $1.72 \pm 0.46$ & $0.36 \pm 0.12$ & $0.45 \pm 0.12$\\
        $13.9 \pm 0.9$ & 19    & 905 & $1.01 \pm 0.29$ & $0.21 \pm 0.07$ & $0.27 \pm 0.08$\\
        \hline
    \end{tabular}
\end{table*}

The cross-correlation between the arrival time of photons at the two detectors is computed in real time by the TDC using exposure times of 10\,s. After the acquisition, the \gtau functions are averaged together after being time shifted from the computed sidereal optical delay between the telescopes~\cite{Guerin:2018}.

\begin{figure}
    \centering\includegraphics{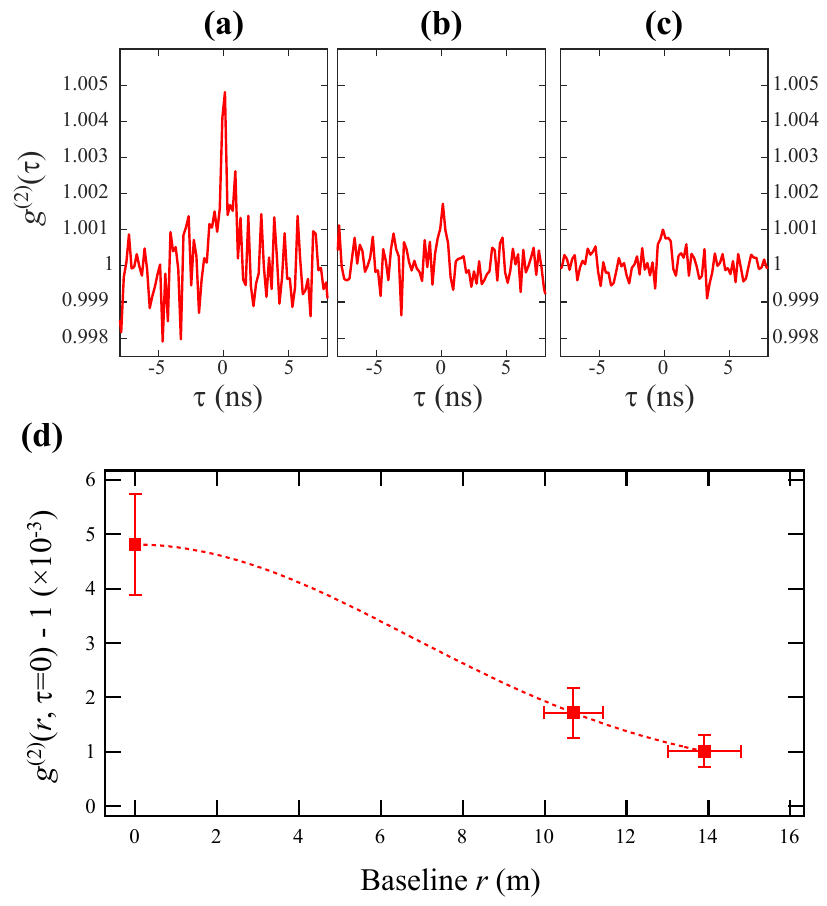}
    \caption{Top row, experimental curves \gtau for different baselines: (a) single telescope experiment\,; (b) two-telescope data for all baselines $r<12$\,m, the average baseline is $r=10.7$\,m\,; (c) the same for $r>12$\,m, average baseline $r = 13.9\,$m. (d) Contrast of the \gtau function as a function of the baseline. The vertical error bars indicate the rms noise and the horizontal error bars the rms width of the distribution of projected baselines during the integration time, neglecting the aperture of the telescopes. The line is only a guide to the eye.}
    \label{fig.g2r}
\end{figure}

Since the projected baseline also changes due to earth rotation, several partial averaging of the data as a function of the computed baseline allows us to obtain several \gtau functions for different projected baselines. Here, the limited signal-to-noise ratio of the data allows us to obtain only two significant curves, for projected baseline $9.5<r<12$\,m and $12<r<15$\, corresponding, respectively, to averaged baselines of 10.7\,m and 13.9\,m. These measurement are reported in Fig.~\ref{fig.g2r}(b,c), along with the single-telescope correlation function [Figs.~\ref{fig.g2tau}(b) and \ref{fig.g2r}(a)] for comparison.

The effect of the partial resolution of \PCyg's emitting envelope is well visible via the contrast of the \gtau function, which is much smaller. This contrast, plotted as a function of the baseline, gives the spatial intensity correlation function $g^{(2)}(r,\tau=0)$, plotted in Fig.~\ref{fig.g2r}(d).
This contrast gives the squared visibility $|V(r)|^2$ [Eq.\,(\ref{eq.Siegert})] after proper normalization, such that $|V(0)|^2 = 1$ at zero baseline. We can use two normalization methods. The most direct method is to divide the contrast measured at $r\neq 0$ by the contrast measured with a single telescope ($r \approx 0$). However the statistical noise as well as any systematic noise due to residual spurious correlations affect the results. The other method is to make use of the measured spectrum and, knowing the temporal resolution of the detection chain, compute the expected zero-baseline contrast, see Fig.\,\ref{fig.spectrum}(d). This method introduces much less noise but relies on the good characterization of the instrumental setup (filter and temporal resolution), which can be done in the lab. We show the results of the two methods in Tab.~\ref{tab.Results} and use the spectrum-based method in Fig.\,\ref{fig.theory} at the end of the next section.

\section{Model of \PCyg\ and comparison with the experiment}\label{sec.Armando}

\subsection{Atmosphere models: code CMFGEN}

In order to analyse the visibility curve of \PCyg, we used state-of-the-art atmosphere models computed with the non-LTE (local thermodynamic equilibrium) radiative transfer code CMFGEN \cite{Hillier:1998}. It solves the coupled problem of the radiative transfer, statistical and radiative equilibrium equations in a spherically symmetric outflow. CMFGEN  has been widely used in the literature to analyse central stars of planetary nebula \cite{Marcolino:2007}, OB-type \cite{Bouret:2012}, LBV \cite{Groh:2012}, Wolf-Rayet stars \cite{Tramper:2013}, and also core-collapse supernovae \cite{Dessart:2016}. It includes, for example, effects of line-blanketing, wind clumping, and Auger ionization by X-rays, thus providing realistic spectra for hot stars from the ultraviolet (UV) to the mid-infrared.\par

The code requires an initial estimate of the hydrostatic structure. For this purpose, we used the BSTAR2006 \cite{Lanz:2007} grid of non-LTE plane-parallel models calculated with the code TLUSTY \cite{Hubeny:1995}. This grid provides pure-photospheric models with effective temperature 15000 K $\leq$ $T_{\mathrm{eff}}$ $\leq$ 30000 K and surface gravity $1.75 \leq \log(g) \leq 3.00$. Up to date, CMFGEN does not allow to calculate hydrodynamically self-consistent models, thus the wind velocity needs to be parameterized. For the wind, we employed a standard $\beta$ velocity law
\begin{equation}
v(r) = v_\infty\left(1 - \frac{R_{\star}}{r}\right)^{\beta},
\end{equation}
where $R_{\star}$ is the stellar radius and $v_\infty$ is the wind terminal velocity. The wind velocity structure is smoothly connected to the hydrostatic structure just above the sonic point.

Clumping was included by default in the models. In CMFGEN, a volume filling factor is used to parameterize the effect of clumping (microclumping approximation) in the wind density structure as follows,
\begin{equation}
f(r) = f_{\infty} + (1 - f_{\infty})\mathrm{e}^{-\frac{v(r)}{v_{\mathrm{initial}}}},
\end{equation}
where  $v_{\mathrm{initial}}$  is the onset velocity of clumping, corresponding to the distance in the wind where inhomogeneity starts to be relevant, and $f_{\infty}$ is the filling factor value at $r\to\infty$. Thus, the density structure is parameterized, including the factor $f(r)$, as follows ($\dot{M}$ is the mass-loss rate):
\begin{equation}
\rho(r) = \frac{\dot{M}}{4\pi r^{2} v(r) f(r)}.
\end{equation}

We did not include Auger ionization by X-rays in the models since \PCyg\ is known to present a very low X-ray luminosity. The X-ray survey on Galactic LBVs of \cite{Naze:2012} could just provide an upper limit of $\log(L_{\textrm{X}}/L_{\textrm{BOL}}) < -9.4$ for \PCyg, including this star in their sub-sample for non-detection of X-ray emission. For comparison, O-type stars typically present $\log(L_{\textrm{X}}/L_{\textrm{BOL}}) \sim -7.0$  \cite{Rauw:2015}.

\subsection{Stellar and wind parameters}

%We analysed CMFGEN models based on the stellar and wind parameters derived by \cite{Najarro:1997} and \cite{Najarro:2001}. Also using CMFGEN, \cite{Najarro:1997} performed a detailed spectroscopic analysis of \PCyg\ in the visible region, while \cite{Najarro:2001} extended their analysis to a multi-wavelength spectroscopic approach from the UV up to the mid-infrared.

CMFGEN is well-suited for analysing \PCyg\ since previous spectroscopic and interferometric studies showed that its wind is almost spherical (see Sect.~\ref{sec.intro}). Following the approach of \cite{Richardson:2013}, we analysed CMFGEN models based on the stellar and wind parameters derived by \cite{Najarro:2001}. Also using CMFGEN, \cite{Najarro:2001} performed a detailed multi-wavelength spectroscopic analysis of \PCyg\ from the ultraviolet up to the mid-infrared region.

%------------------------------------------------------
\begin{table}
\caption{\label{atomic_species} Number of levels, super-levels, and bound-bound transitions for each atomic species included in our CMFGEN reference model.}
\centering
\renewcommand{\arraystretch}{1.1}
%\begin{adjustbox}{width=0.60\textwidth}
\begin{tabular}{lcccc}
\toprule
\toprule
Ion & Full-levels & Super-levels & b-b transitions \\
\midrule
\ion{H}{I} & 30 & 30  & 435\\

\ion{He}{I} & 69 & 69 & 905 \\
 \ion{He}{II} & 30 & 30 & 435 \\

\ion{C}{II} & 100 & 44 & 1064 \\
\ion{C}{III}  & 99 & 99 & 5528 \\
\ion{C}{IV}  & 64 & 64 & 1446 \\

\ion{N}{I} & 104 & 44 & 855 \\
\ion{N}{II} & 144 & 62 & 1401 \\
\ion{N}{III} & 287 & 57 & 6223 \\

\ion{O}{I} & 90 & 35 & 615 \\
\ion{O}{II} & 123 & 54 & 1375\\
\ion{O}{III} & 104 & 36 & 761 \\

\ion{Mg}{II} & 44 & 36 & 348 \\

\ion{Al}{II} & 44 & 26 & 171 \\
\ion{Al}{III} & 65 & 21 & 1452 \\

\ion{Si}{II} & 62  & 34 & 365 \\
\ion{Si}{III} & 50  & 50 & 232 \\
\ion{Si}{IV} & 66  & 66 & 1090 \\

\ion{S}{II} & 88 & 27 & 796 \\
\ion{S}{III} & 41 & 21 & 177 \\
\ion{S}{IV} & 92 & 37 & 708 \\

\ion{Ca}{II} & 19 & 12 & 65 \\

\ion{Fe}{II} & 510 & 111 & 7357 \\
\ion{Fe}{III} & 607 & 65 & 5482\\
\ion{Fe}{IV} & 1000 & 100 & 25241 \\
\ion{Fe}{V} & 1000 & 139 & 25173 \\
\bottomrule
\end{tabular}
%\end{adjustbox}
\end{table}
%------------------------------------------------------

In Table \ref{atomic_species}, we show the atomic species included in the models together with the number of energy levels\footnote{Super-level approach (grouping of energy levels) is introduced in CMFGEN for a faster computational treatment. See \cite{Hillier:1998} for further details.} and bound-bound transitions. These model atoms are similar to those used by \cite{Najarro:2001}, providing a rather robust model to reproduce the spectrum of \PCyg\ in the UV, visible, and infrared regions. We also assumed the same chemical abundances as \cite{Najarro:2001}. Since \PCyg\ has ended the hydrogen core-burning phase \cite{Langer:1994}, the assumption of solar chemical abundances ($Z_{\odot}$) must overestimate the intensity in the H$\alpha$ line (considering a fixed set of physical parameters in the model). Most important for the comparison with the observed visible spectrum, the abundances of H, He, C, N, O were set to 0.66, 1.86, 0.31, 6.5, and 0.18 $Z_{\odot}$, respectively.\par

%---------------------------------------%---------------------------------------

%---------------------------------------%---------------------------------------

In Table \ref{parameters_cmfgen}, we present the physical stellar and wind parameters of our CMFGEN reference model. These are the main parameters to define the atmosphere model: stellar luminosity ($L_{\star}$), effective temperature ($T_{\mathrm{eff}}$), gravity surface acceleration ($\log g$), radius ($R_{\star}$), mass ($M_{\star}$), mass-loss rate ($\dot{M}$), wind clumping factor ($f_{\infty}$), terminal velocity ($v_{\infty}$), and the wind velocity law exponent ($\beta$). Except for the surface gravity $\log g$, all the other parameters are equal or close to the ones derived by \cite{Najarro:2001}. We set $\beta$ = 2.3 in our reference model due to numerical issues with $\beta$ = 2.5 \cite{Najarro:2001}. As will be discussed in Sect. \ref{sec_comp_spectroscopic_data}, we set $\dot{M}$ = $4.0\e{-5}$ $\mathrm{M_\odot}$ yr\textsuperscript{-1} instead of $\dot{M}$ = $2.4\e{-5}$ $\mathrm{M_\odot}$ yr\textsuperscript{-1} \cite{Najarro:2001}. Instead of $\log g$ = 1.20, as in \cite{Najarro:2001}, we assumed $\log g$ = 2.25 since this is the lower value of $\log g$ in the TLUSTY models, according to the used effective temperature ($T_{\mathrm{eff}}$ = 18700 K). Nevertheless, as pointed out by \cite{deJager:2001}, the determination of this parameter for \PCyg\ is quite uncertain, with a discrepancy up to a factor of 10 from different works in the literature. For example, \cite{Pauldrach:1990} derived 2.04 for the surface gravity of \PCyg.\par

\begin{table}
\caption{\label{parameters_cmfgen} Summary of the main stellar and wind parameters of our CMFGEN reference model.}
\centering
\renewcommand{\arraystretch}{1.2}
%\begin{adjustbox}{width=0.36\textwidth}
\begin{tabular}{ll}
\toprule
\toprule
%\multicolumn{1}{c|}{ Continuum (635-650nm)} \\
%\midrule
$L_{\star}$ ($\mathrm{L_{\odot}}$) & 610000  \\
$T_{\mathrm{eff}}$ (K) & 18700 \\
$\log g$ & 2.25 \\
$R_{\star}$ ($\mathrm{R_{\odot}}$) & 75  \\
$M_{\star}$ ($\mathrm{M_{\odot}}$) & 37  \\
\midrule
$\dot{M}$ ($\mathrm{M_\odot}$ yr\textsuperscript{-1}) & $4.0\e{-5}$ \\
$f_{\infty}$ & 0.5 \\
$v_{\infty}$ (km s\textsuperscript{-1}) & 185 \\
$\beta$ & 2.3 \\
\bottomrule
\end{tabular}
%\end{adjustbox}
\end{table}

\subsection{Results of the simulations}

\subsubsection{Comparison to spectroscopic data}
\label{sec_comp_spectroscopic_data}

\begin{figure}[t]
%\centerline{\resizebox{1.20\textwidth}{!}{\includegraphics[angle=0]{./pcygni_aras_data_model_e28.pdf}}}
\centering\includegraphics[width=\columnwidth]{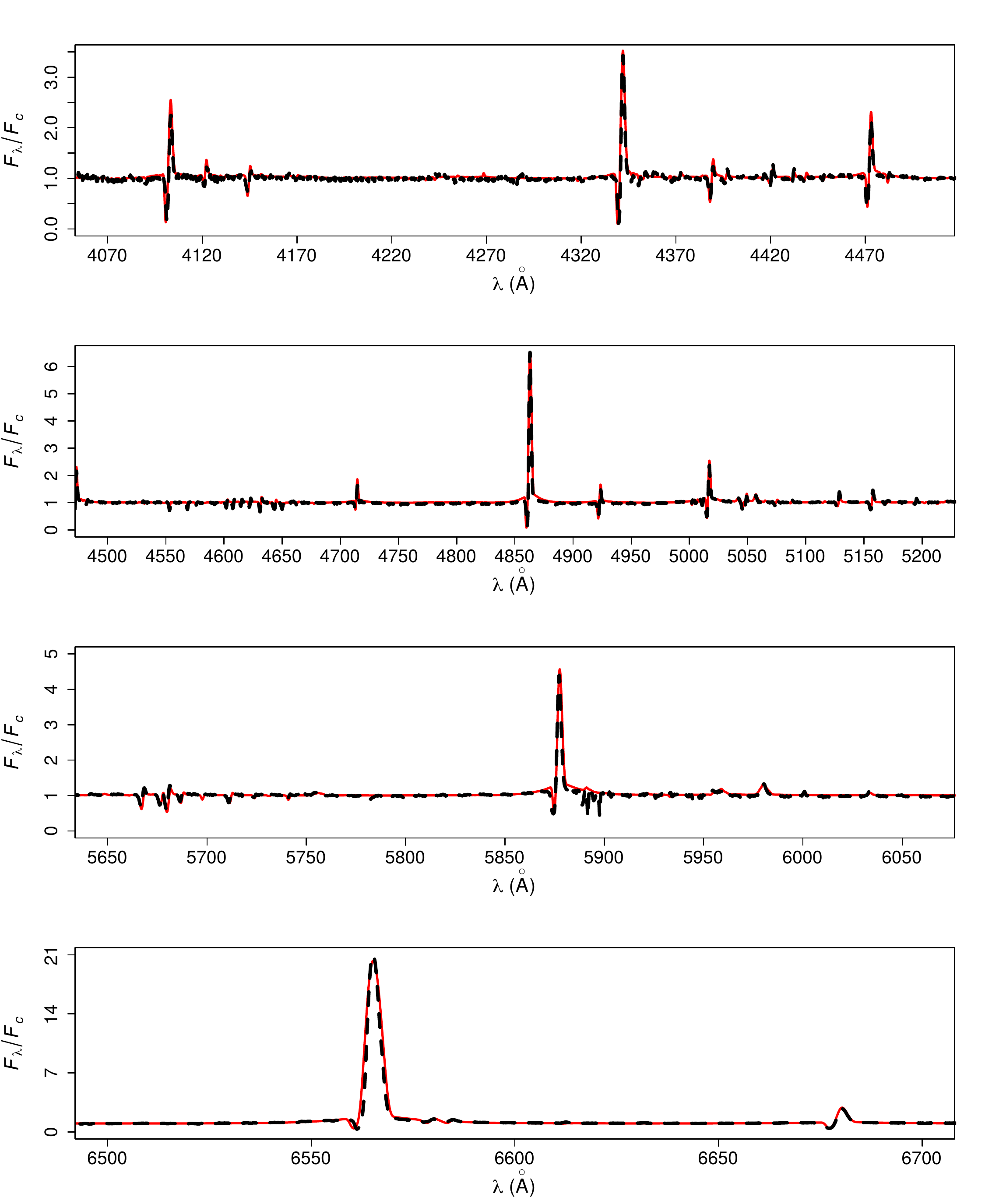}
\caption{Comparison between the observed spectrum of \PCyg\ (black dashes) and the spectrum of the CMFGEN reference model (red line) used to analyze the interferometric data. The mass-loss rate of our reference model ($\dot{M}$ = $4.0\e{-5}$ $\mathrm{M_\odot}$ yr\textsuperscript{-1}) is close to the one derived by \cite{Najarro:2001} of $\dot{M}$ = $2.4\e{-5}$ $\mathrm{M_\odot}$ yr\textsuperscript{-1}. This model provides a fairly reasonable overall match to the spectrum.}
\label{comparison_spectrum_aras_data_cmfgen}
\end{figure}

Before analyzing our interferometric data, we compare, in Fig. \ref{comparison_spectrum_aras_data_cmfgen}, the synthetic spectrum calculated from our CMFGEN reference model (Tables\,\ref{atomic_species} and \ref{parameters_cmfgen}) to the observed spectrum of \PCyg\ in the visible region, obtained from the ARAS Spectral Data Base. This comparison allows a physical validation, in terms of the spectroscopic appearance, of our adopted atmosphere model. Due to the effect of radial velocity, the observed spectrum was shifted in wavelength in order to match the synthetic spectrum.\par

Fig.\,\ref{comparison_spectrum_aras_data_cmfgen} shows that our reference model is able to reproduce well the observed visible spectrum of \PCyg, showing intense P Cygni profiles in the Balmer and helium lines. Overall, the weak spectral features due to metals, such as \ion{C}{II} $\lambda$6580 and $\lambda$6585 (close to H$\alpha$), are also fairly reproduced. Initially, we assumed the same value for the mass-loss rate as \cite{Najarro:2001}, i.e., $\dot{M}$ = $2.4\e{-5}$ $\mathrm{M_\odot}$ yr\textsuperscript{-1} with $f_{\infty}$ = 0.5. Since the emission component of H$\alpha$ is highly sensitive to the variation of the mass-loss rate, we followed the simplest approach of only varying this fundamental parameter of the wind. The Balmer lines, in particular H$\alpha$, seems to be better reproduced using a slightly higher value for the mass-loss rate ($\dot{M}$ = $4.0\e{-5}$ $\mathrm{M_\odot}$ yr\textsuperscript{-1} with $f_{\infty}$ = 0.5). This difference is encompassed by the typically uncertainties on $\dot{M}$ found from spectroscopic analysis of massive stars in literature \cite[see, e.g.,][]{deAlmeida:2019}. In addition, it is very unlikely to have a physical cause, as this difference is much larger than the mass-loss rate variability of \PCyg\ \cite{Markova:2001}.

It is beyond the scope of this paper to derive the stellar and wind parameters of \PCyg, as performed by \cite{Najarro:2001}. Nevertheless, the ability of our CMFGEN reference model to reproduce the visible spectroscopic appearance of \PCyg\ makes us confident to adopt this model in order to interpret our II observed visibilities.

\subsubsection{Comparison to normalized II visibilities}\label{sec.visibilities}

\begin{figure}
\centering\includegraphics[width=\columnwidth]{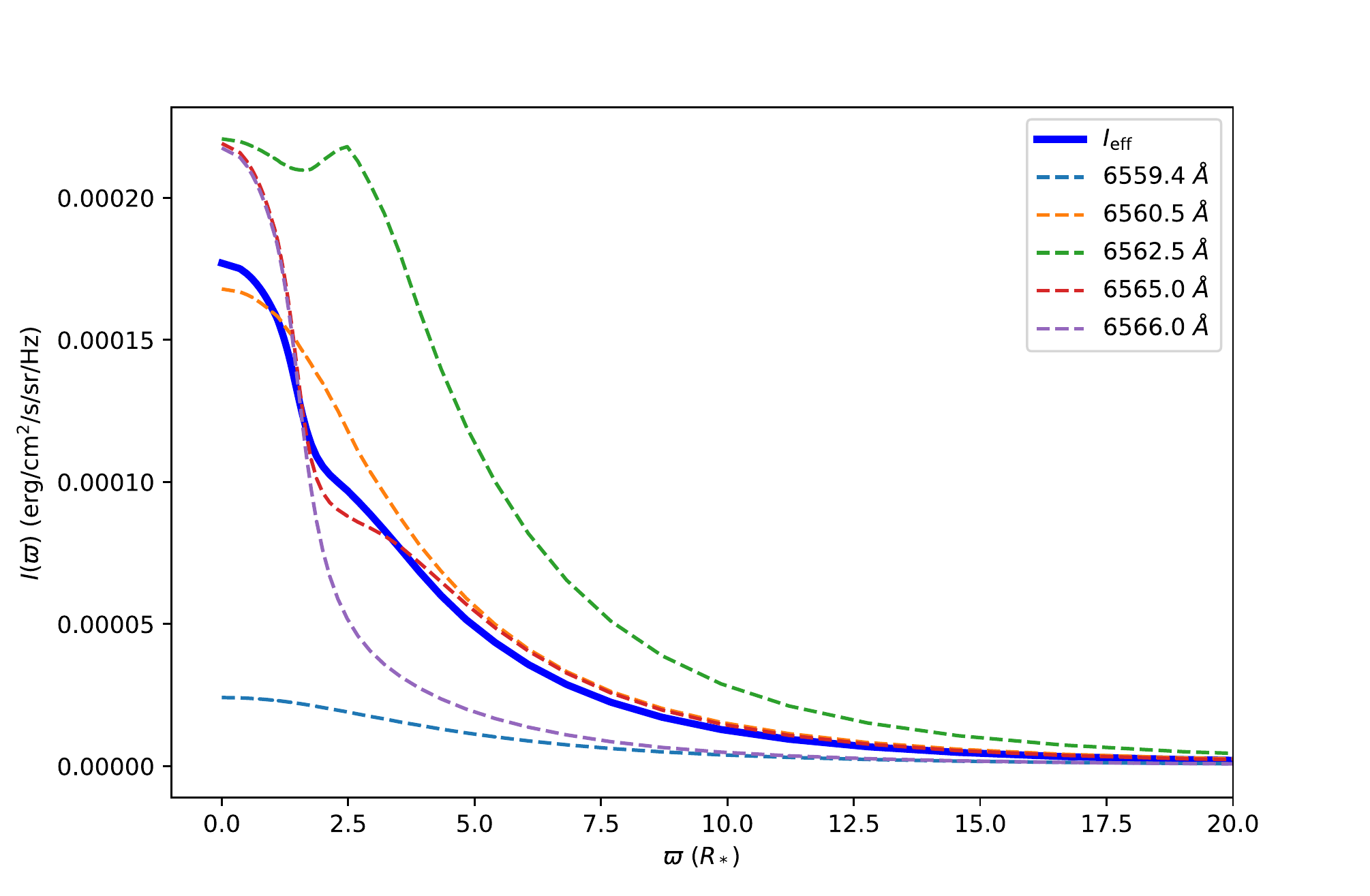}
\caption{Effective H$\alpha$ radial profile $I_\mathrm{eff}(\varpi)$ (Eq.~\ref{eq_Ieff}) of the reference CMFGEN model (thick solid blue) as a function of the radial coordinate $\varpi$ given in units of the stellar photospheric radius (clipped at $20R_*$ for better visualization). For comparison, the dashed curves show the model specific intensity profiles $I(\lambda,\varpi)$ for selected wavelengths within the H$\alpha$ emission line and in the region where the filter transmission is high. In particular, we show the profile at $\lambda=6562.5$\,\AA, nearly at the maximum of the model H$\alpha$ spectrum. We note that these selected profiles were not multiplied by the filter transmission.}
\label{fig.effective_radial_profile_cmfgen}
\end{figure}

To compare the reference CMFGEN model of \PCyg\ to the normalized II visibilities we need to compute the effective radial intensity profile $I_\mathrm{eff}(\varpi)$ corresponding to the observed spectral region within the H$\alpha$ filter,
\begin{equation} \label{eq_Ieff}
%See Bessell1998_v333p231-250 (A.1)
I_\mathrm{eff}(\varpi) = \frac{\int I(\lambda,\varpi) T(\lambda) \dd \lambda}{\int T(\lambda) \dd \lambda} \, ,
\end{equation}
where $I(\lambda,\varpi)$ is the 1D monochromatic specific intensity, provided by CMFGEN, as a function of the radial coordinate $\varpi$ (impact parameter). The effective wavelength $\lambda_\mathrm{eff}$ corresponding to $I_\mathrm{eff}$ for the reference CMFGEN model is given by
\begin{equation}
\lambda_\mathrm{eff} = \frac{\int \lambda F(\lambda) T(\lambda) \dd \lambda}
            {\int F(\lambda) T(\lambda) \dd \lambda}=6562.9\, \text{\AA} \, .
\end{equation}
As before, $F(\lambda)$ and $T(\lambda)$ are the observed spectrum and the transmission filter, respectively (see Fig.~\ref{fig.spectrum}). The effective H$\alpha$ radial profile $I_\mathrm{eff}(\varpi)$ of the reference CMFGEN model is shown in Fig.~\ref{fig.effective_radial_profile_cmfgen}, together with radial profiles at some selected wavelengths for comparison.

\begin{figure}
%\centering\includegraphics[width=\columnwidth]{V2_vs_Bproj_e28_eff_lam_shift_best-fit_theo_norm.pdf}
\centering\includegraphics[width=\columnwidth]{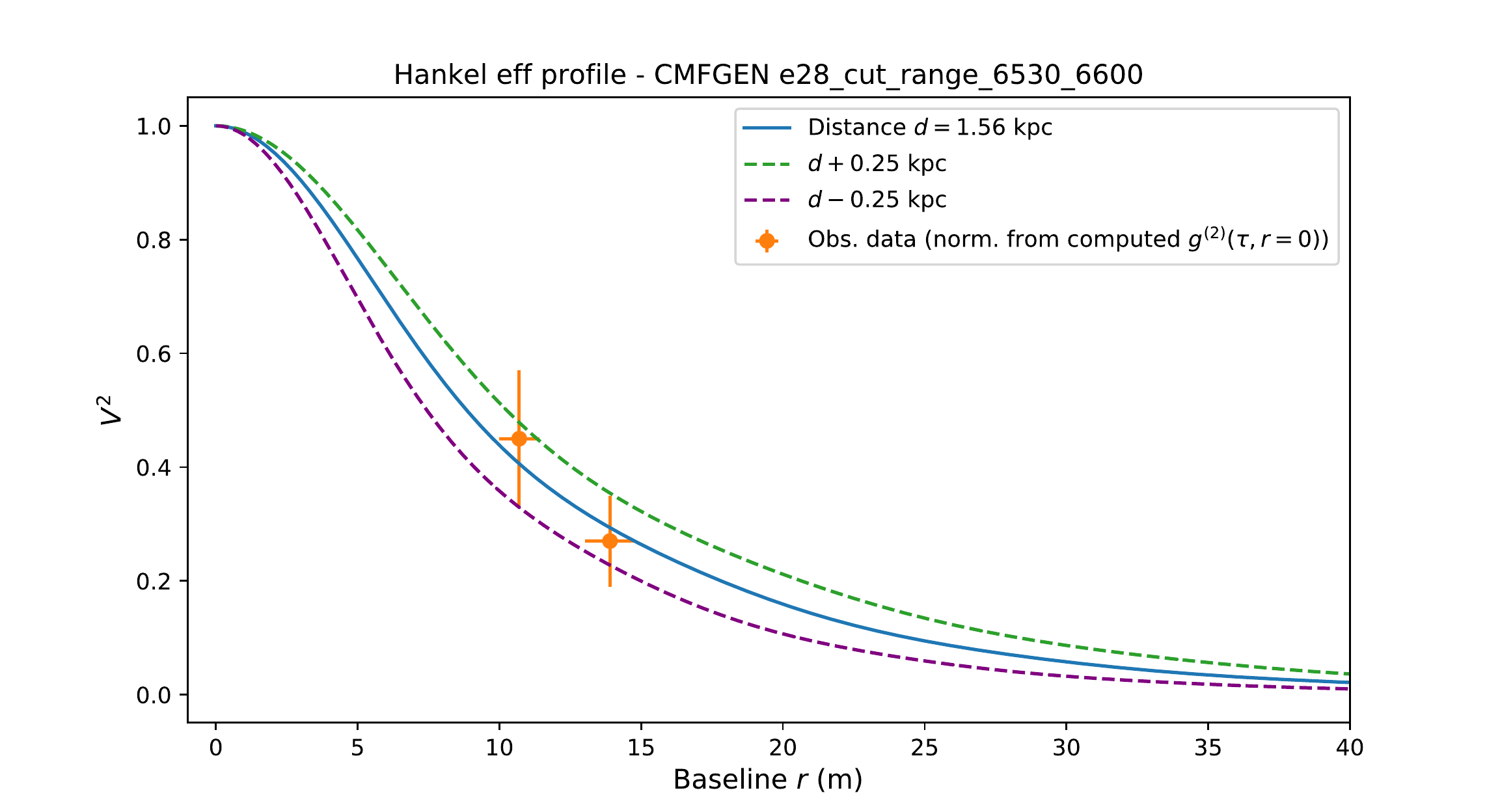}
\caption{The data points are the measured squared visibility (section~\ref{sec.two-telescope}) normalized by the contrast computed from the measured spectrum (Fig.~\ref{fig.spectrum}d). They are fitted (solid blue line) using Eq.\,(\ref{v2_hankel_Ieff}) with the distance $d$ to \PCyg\ as the only free parameter (further details in the text). The curves correspond to the best-fit $d$ (solid) and associated $\pm1\sigma$ uncertainties (dashed).
}
\label{fig.theory}
\end{figure}

The normalized squared visibility $|V|^2$ (or simply $V^2$) associated to the reference CMFGEN model is computed thanks to the Hankel transform of $I_\mathrm{eff}(\varpi)$, normalized by the corresponding spectral flux, as
\begin{equation} \label{v2_hankel_Ieff}
V^2=\left| \frac{\int_{0}^\infty I_\mathrm{eff}(\rho) J_0(2\pi\rho q) 2\pi\rho \dd \rho}
    {\int_{0}^\infty I_\mathrm{eff}(\rho) 2\pi\rho \dd \rho} \right|^2 \, ,
\end{equation}
where $J_0$ is the zeroth-order Bessel function of the first kind, $\rho=\varpi/d$ is the radial angular coordinate, with $d$ being the distance to the target. The radial spatial frequency coordinate associated to $\rho$ is $q=r/\lambda_\mathrm{eff}$, i.e. the II average projected baseline $r$ divided by the effective wavelength of the observations $\lambda_\mathrm{eff}$. The Hankel transform is used here because of the circular symmetry of the model.

To interpret the II observations we used the above equations to compute the $V^2$ corresponding to our reference CMFGEN model, which we assume to be a \textit{bona fide} representation of \PCyg, since it well reproduces the observed visible spectrum, as shown in the previous subsection. Under this assumption, the only remaining free parameter is the distance $d$. We have thus used a Python-Scipy non-linear least squares routine to fit the reference model $V^2$ to our II data, which allowed us to estimate the distance to \PCyg\ as $d=1.56\pm0.25$~kpc. The fit has been performed on the visibility data normalized by the zero-baseline visibility computed from the measured spectrum (Fig.\,\ref{fig.spectrum}d and Table\,\ref{tab.Results}), as those data are less subjected to spurious correlations than the measured single-telescope correlation function. The latter has thus not been used.
The observed and best-fit model $V^2$ are shown in Fig.\,\ref{fig.theory}. These results and the interpretation of the measured $d$ are discussed in the following section.

%%%%%%%%%%%%%%%%%%%%%%%%%%%%%%%%%%%%%%%%%%%%%%%%%%%%%%%%%%%%%
\section{Discussion and conclusion}\label{discussion}

%\PCyg, together with $\eta$ Car, are the brightest and most studied LBV stars for their spectrometric and photometric observational aspects, which have largely served to determine  their physical properties \cite{Najarro:2001}. More recently, high angular resolution data, especially from long-baseline interferometry, have shed a new light on the fine spatial details of their mass loss and geometries in general\,\cite{Weigelt:2007}. As for \PCyg, GI2T, NPOI and CHARA interferometers (see our introduction section) provide valuable estimates of the star parameters and the morphology of its wind at the level of mas or a few mas angular resolutions. Although their results agree qualitatively, the accurate numbers differ on the extent of the H$\alpha$ emitting envelope for instance. It has been suggested that these differences were due to the variability of \PCyg\ on time scales of a few months to years, as confirmed by photometric and spectrometric monitoring of \PCyg\ in H$\alpha$ and H band IR observations. Richardson \textit{et al.}~\cite{Richardson:2013} have also carried aperture synthesis imaging at two different epochs in 2010 and 2011 with the CHARA array and the combination of these data sets suggest some marginal asymmetry, although no evidence was found to confirm or not the wind asymmetry of \PCyg\ found by the GI2T based on differential phase measures~\cite{Vakili:1997}.

% New shorter version
\PCyg, together with $\eta$ Car, are the brightest and most studied LBV stars for their spectrometric and photometric observational aspects, which have largely served to determine  their physical properties \cite{Najarro:2001}. More recently, high angular resolution data, especially from long-baseline interferometry, have shed a new light on the fine spatial details of their mass loss and geometries in general\,\cite{Weigelt:2007}. As for \PCyg, GI2T, NPOI and CHARA interferometers provided valuable estimates of the star parameters at the level of mas or a few mas angular resolutions. Although their results agree qualitatively, the numbers differ on the extent of the H$\alpha$ emitting envelope, for instance, which might be due to the variability of \PCyg\ on time scales of a few months to years.

The method to determine the extent of \PCyg\ has often used analytical models such as uniform, limb-darkened disks or multiple Gaussians, whilst authors adopt distance values from different techniques, e.g. O-B association membership~\cite{Lamers:1983,Turner:2001}, to interpret measured visibility points. The distance controversy for \PCyg\ is well known, with distances determined from $\sim 1.2$\,kpc up to 2.3\,kpc \cite[see, e.g., Table 1 of][]{Turner:2001}. For instance CHARA studies~\cite{Richardson:2013} adopted a 1.7~kpc distance of \PCyg\ to match synthetic visibilities based on the CMFGEN stellar atmospheric model and basic parameters~\cite{Najarro:2001} to their observed visibilities. The more accurate distance for \PCyg, $d_\mathrm{G}=1.36\pm0.24$~kpc, from the Gaia global astrometry mission and its second data release DR2~\cite{Gaia:2018}, could also be used. However Gaia has been designed for sources fainter than 11th magnitude in the visible, where the parallax determination is limited by the photon noise. \PCyg\ has a visual magnitude of 4.5 which is too bright for the normal scanning operation at the focal detector of Gaia (F.\,Mignard, private communication). Therefore the question of the exact parallax of \PCyg\ remains a real issue.

In this context, we have followed a different route by fixing the linear size of \PCyg\ in agreement with detailed multi-wavelength spectroscopic studies in the literature (see section\,\ref{sec.Armando}). Our adopted model reproduces fairly well several lines of different atomic species, allowing us to adopt the linear radius of \PCyg\ photosphere as 75 $R_{\odot}$ and deliver synthesized visibilities and finally determine the distance of \PCyg\ as $d_\mathrm{II}=1.56\pm 0.25$\,kpc. Note that for such an approach to be effective, it would be useful to monitor interferometric measurements by simultaneous (and preferably U, B, V) photometry. For \PCyg\ the variability can originate from effective temperature and radius changes of 10\% and 7\% respectively, as concluded by~\cite{Markova:2001}.

With this rather unusual interpretation of long baseline interferometry data we propose a method to check and improve the so-called \textit{Wind Momentum versus Luminosity relation} (W-LR hereafter) introduced by \cite{Kudritzki:1994}, which relates the momentum flow of the wind from the star to its linear size times its luminosity,
\begin{equation}\label{eq.WLR}
\dot{M} v_\infty \propto R_{\star}^{-1/2}L_{\star}^{-1/\alpha_\mathrm{eff}},
\end{equation}
where $\alpha_\mathrm{eff}$ reflects all the spectral lines that drive the wind, with a typical value of 2/3, varying according to the spectral-type \cite[see, e.g., Table 2 of ][]{Kudritzki:2000}.

The W-LR method consists in recording medium- or high-resolution spectra of the most luminous stars such as O, B, A supergiants, B[e] and LBV stars of the nearby galaxies, or the local Universe if possible, and determine their intrinsic luminosity from quantitative spectroscopy. Despite the good agreement between the theoretical and empirical (derived from spectroscopic analyses) W-LR for the most luminous massive stars (as the ones mentioned above), O-type dwarfs and giants with $\log L_{\star}/L_{\odot}$ $\leq$ 5.2 present much lower values of mass-loss rate, up to 2 orders of magnitude, than the theoretical values, affecting the WL-R \cite{Martins:2005,Marcolino:2009,deAlmeida:2019}. This shows the current need to check independently the WL-R.

To our knowledge, CMFGEN radiative transfer code represents a robust model to carry such a quantitative spectroscopy of the most luminous stars with their emission lines that often possess P Cyg profiles, i.e. a strong emission red wing and a blue absorption component corresponding to the projection of the wind components on the line of sight. The comparison of the apparent magnitude to the absolute luminosity would then estimate the distance of the luminous star to us even at Megaparsec levels.

As suggested by \cite{Vakili:1998}, such a method could be further improved by carrying the quantitative spectroscopy of a star observed by long-baseline interferometry and matching synthesized visibilities based on linear diameter of the star versus the measured visibility so as to determine the star's distance. This approach could be furthermore improved by observing luminous stars of Magellanic Clouds with different chemical abundances, i.e. LMC versus SMC. The brightest stellar members of Magellanic Clouds have apparent magnitudes in the range of 12 to 15 in the visible and their visibilities could be measured with future extremely long-baseline optical interferometers such as the intensity interferometric mode of the CTA array \cite{Dravins:2016} or connecting large optical telescopes on existing observatories~\cite{Lai:2018}, such as Mauna Kea or Paranal, which will offer better than 10 $\mu$as angular resolution, compatible with the range of angular diameters of the brightest stars of the Magellanic Clouds. Therefore the present work constitutes the first successful step towards settling the quantitative spectroscopy of luminous stars and the W-LR relation, which may serve as an independent calibration technique of cosmological distances comparable to the Cepheid or post-AGB methods\,\cite{Whitelock:2012}.

%%%%%%%%%%%%%%%%%%%%%%%%%%%%%%%%%%%%%%%%%%%%%%%%%%%%%%%%%%%%%
\section*{Acknowledgements}

We thank P.~Weiss for participating in the observations, G.~Labeyrie for fruitful discussions, and the M\'eO Team (G\'eoazur Laboratory) for their hospitality and help during the tentative session in November 2018. We also thank F.~Th\'evenin, P.~B\'erio and the Observatoire de la C{\^o}te d'Azur for financial support. We acknowledge the CATS team (Calern Atmospheric Turbulence Station, https://cats.oca.eu/) for providing turbulence real-time measurements and Swabian Instruments for very reactive technical support. We are also grateful to the ARAS database community members, especially to J. Guarro i Fl{\^o}, for their high quality spectra of \PCyg. This work is supported by the UCA-JEDI project ANR-15-IDEX-01, the Doeblin Federation and the OPTIMAL platform. E. S. G. de Almeida thanks OCA and the ``Ville de Nice'' (Nice, France) for the financial support to this work through the ``Bourse Doctorale Olivier Chesneau'' during the period of 2016-2019.


%\bibliographystyle{apsrev4-2}
%\bibliography{HBT_paper_biblio}


%%%%%%%%%%%%%%%%%%%%%%%%%%%%%%%%%%%%%%%%%%%%%%%%%%

\end{document}